\documentclass[sigconf]{acmart}

\usepackage{hyperref}
\usepackage{graphicx}
\usepackage{booktabs}
\usepackage{caption}
\usepackage{subcaption}
\usepackage{multirow}
\usepackage{enumitem}
\usepackage{bbm}
\usepackage{xspace}
\usepackage[ruled,linesnumbered]{algorithm2e}
\usepackage{amsmath}
\usepackage{braket}
\usepackage{bm}
\usepackage{color,xcolor}
\usepackage[misc]{ifsym}
\usepackage[normalem]{ulem}
\usepackage{bbding}
\usepackage{url}
\useunder{\uline}{\ul}{}
\usepackage[switch]{lineno}

\makeatletter
\newsavebox\myboxA
\newsavebox\myboxB
\newlength\mylenA

\newcommand*\xoverline[2][0.75]{%
\sbox{\myboxA}{$\m@th#2$}%
\setbox\myboxB\null
\ht\myboxB=\ht\myboxA%
\dp\myboxB=\dp\myboxA%
\wd\myboxB=#1\wd\myboxA
\sbox\myboxB{$\m@th\overline{\copy\myboxB}$}
\setlength\mylenA{\the\wd\myboxA}
\addtolength\mylenA{-\the\wd\myboxB}%
\ifdim\wd\myboxB<\wd\myboxA%
\rlap{\hskip 0.5\mylenA\usebox\myboxB}{\usebox\myboxA}%
\else
\hskip -0.5\mylenA\rlap{\usebox\myboxA}{\hskip 0.5\mylenA\usebox\myboxB}%
\fi}
\makeatother

\newcommand{\ie}{\emph{i.e.,}\xspace}
\newcommand{\eg}{\emph{e.g.,}\xspace}

\newcommand{\etal}{\emph{et al.}\xspace}
\newcommand{\paratitle}[1]{\vspace{1.5ex}\noindent\textbf{#1}}
\newcommand{\wrt}{\emph{w.r.t.}\xspace}
\newcommand{\ignore}[1]{}

\newcommand{\our}{TedRec}

\AtBeginDocument{%
  \providecommand\BibTeX{{%
    \normalfont B\kern-0.5em{\scshape i\kern-0.25em b}\kern-0.8em\TeX}}}

\acmConference[Conference acronym'XX]{}{June 03–05,
  2018}{Woodstock, NY}
\begin{document}

\title{Sequence-level Semantic Representation Fusion for Recommender Systems}
\author{Lanling Xu$^{1}$, Zhen Tian$^{1}$, Bingqian Li$^{1}$, Junjie Zhang$^{1}$, Jinpeng Wang$^{2}$ \\ Mingchen Cai$^{2}$, Wayne Xin Zhao$^{1}$\textsuperscript{\Letter}}
\thanks{\textsuperscript{\Letter} Wayne Xin Zhao (batmanfly@gmail.com) is the corresponding author.}

\affiliation{%
  \institution{$^1$Gaoling School of Artificial Intelligence, Renmin University of China, China}
  \institution{$^2$Meituan Group, Beijing China}
  \country{}
}

\renewcommand{\shortauthors}{Lanling Xu et al.}

\renewcommand{\shortauthors}{Anonymous}

\begin{abstract}
With the rapid development of recommender systems, there is increasing side information that can be employed to improve the recommendation performance.
Specially, we focus on the utilization of the associated \emph{textual data} of items (\eg product title) and study how text features can be effectively fused with ID features in sequential recommendation. 
However, there exists distinct data characteristics for the two kinds of item features, making a direct fusion method (\eg adding text and ID embeddings as item representation) become less effective.  
To address this issue, we propose a novel {\ul \emph{Te}}xt-I{\ul \emph{D}} semantic fusion approach for sequential {\ul \emph{Rec}}ommendation, namely \textbf{\our}. 
The core idea of our approach is to conduct a sequence-level semantic fusion approach by better integrating global contexts. The key strategy lies in that we transform the text
 embeddings and ID embeddings by Fourier Transform from  \emph{time domain} to \emph{frequency domain}. In the frequency domain, the global sequential characteristics of the original sequences are inherently aggregated into the transformed representations, so that we can employ simple multiplicative operations to effectively fuse the two kinds of item features. Our fusion approach can be proved to have the same effects of contextual convolution, so as to achieving sequence-level semantic fusion. In order to further improve the fusion performance, we propose to enhance the discriminability of the text embeddings from the text encoder, by adaptively injecting
positional information via a mixture-of-experts~(MoE) modulation method. 
{Extensive experiments on five public datasets demonstrate the effectiveness of our approach by comparing with a number of competitive baselines.} 
Our implementation is available at this repository: \textcolor{magenta}{\url{https://github.com/RUCAIBox/TedRec}}.
\end{abstract}





\maketitle

\section{Introduction}
\label{sec:intro}

Sequential recommender systems, aiming to predict the next interaction items of a user based on her/his historical records, have been a widely studied task in both academia and industry~\cite{li2017narm,hou2022unisrec,du2023fearec}. 
The key to deliver satisfied recommendation service lies in effectively modeling the interaction behavior of users, thus making  accurate prediction of future interactions. 
Therefore, various model architectures including CNN~\cite{tang2018caser}, RNN~\cite{tan2016gru4rec} and Transformer~\cite{kang2018sasrec,sun2019bert4rec} have been developed to characterize sequential patterns of users.

In existing literature, most of sequential recommendation models~\cite{sun2019bert4rec,fan2021lightsans} rely on pure \emph{item identifiers~(IDs)} to model the user behavior.
Despite the effectiveness, 
 ID-only modeling paradigm restricts the effective utilization of extensive context data (\eg item features) in real-world scenarios.
Considering this limitation, a number of approaches~\cite{xi2023towards,liu2023once,yuan2023morec} have been proposed to incorporate the side information into recommender systems for better modeling the actual preference of users.
Specially, the textual data associated with items (\eg item title and category label) have been widely explored for enhancing the performance of sequential recommender~\cite{zhang2019fdsa,zhou2020s3rec,gao2023smlp4rec},  which provide important description information about the characteristics or functions of items. 
\ignore{Recently, considering the excellent capabilities of Large Language Models~(LLMs) in modeling the textual data, a number of approaches~\cite{xi2023towards,liu2023once,yuan2023morec} have been proposed to incorporate the world knowledge of LLMs into recommender systems for better modeling the actual preference of users.
Typically, they leverage LLMs as the text encoder for obtaining its vector representation~(\ie textual embedding), which are further integrated with ID features to serve as the final representations of recommended items.}

\begin{figure}[t]
    \centering
    \captionsetup{font={small}}
    \includegraphics[width=0.45\textwidth]{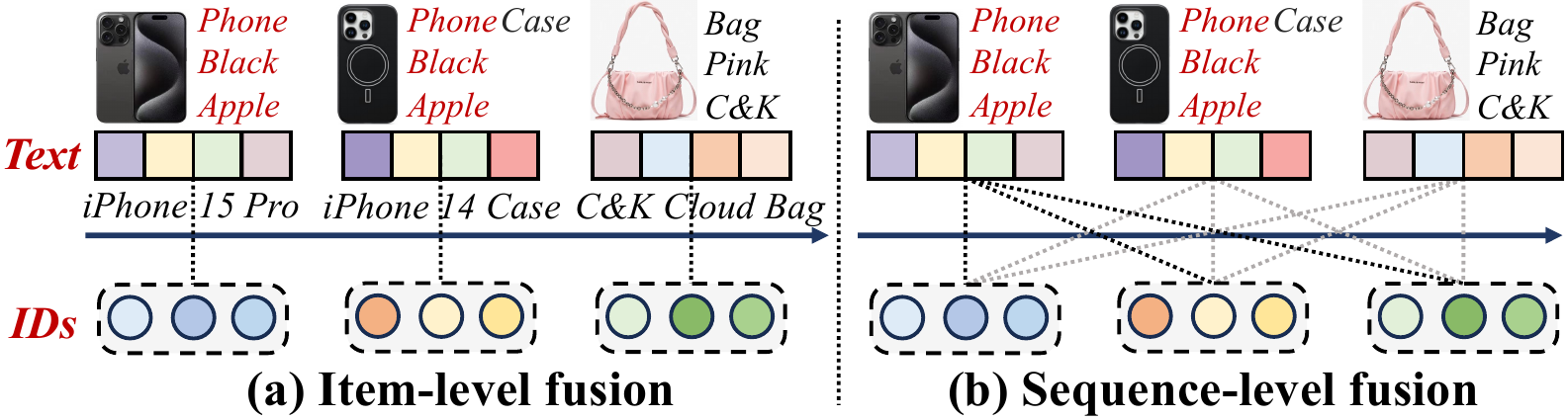}
    \caption{Illustration for two kinds of fusion paradigms.}
    \label{fig:intro_figure}
    \vspace{-1em}
\end{figure}

Typically, these approaches often adopt a relatively simple representation fusion approach, \eg adding the ID embedding and text embedding of an item~\cite{hou2022unisrec}. 
Despite the performance improvement, such a semantic fusion way is locally \emph{at the item level}. More specifically, the auxiliary text embedding is only used to enhance the representation of the target item, which cannot be directly utilized by other positions in the same sequence before attention interaction. In contrast, we argue that a more ideal fusion way should be globally \emph{at the sequence level}: the text embedding at some position can be fused into the global semantics of the entire sequence. We present a comparison between the \emph{item-level} and \emph{sequence-level} semantic fusion in Figure~\ref{fig:intro_figure}. 
As we can see,  sequential-level semantic fusion can better leverage the auxiliary text information of items for improving sequential user behavior modeling. 
Another limitation of existing approaches is that they often employ a pre-trained language model as the text encoder (\eg BERT~\cite{devlin2018bert}) to obtain the text embedding. It has been found that the learned embeddings are less discriminative among {similar items~\cite{hou2022unisrec,li2020sentence,huang2021whiteningbert}.}   
For example, given two item titles ``\emph{iPhone 15 Pro}'' and ``\emph{iPhone 14 case}'', existing text encoders tend to produce very similar representations,  lacking fine-grained discriminatility in similar or related products.


In order to achieve  sequence-level semantic fusion, we are inspired by recent work on the use of \emph{Fourier Transform} in recommender systems~\cite{zhou2022fmlp-rec,du2023fearec,liu2023dlfs-rec}. The key idea of our approach is to leverage the Fourier Transform for transforming the original representations (\ie sequences of text embeddings and ID embeddings) from the original \emph{time domain} to the \emph{frequency domain}.
In this way, the global sequential characteristics of the original sequences can be inherently aggregated into the transformed representations in the frequency domain.  
Further, we can conduct the multiplication fusion  between the transformed textual and ID representations. It can be proved (see Section~\ref{sec:method-theoretical-analysis}) that   
the multiplicative operations (\eg \emph{Hadamard product}) in the frequency domain well align with the sequence-level operations (\eg \emph{convolution}) in the time domain, thus enabling the sequence-level semantic fusion between textual and ID representations. To develop our approach, we highlight two key challenges to be solved: (i) how to effectively fuse the textual representations and ID features in the frequency domain; (ii) how to further enhance the discriminability of text embeddings  for adapting to the recommendation scenarios.  

To this end, in this paper,  
we propose a novel {\ul \emph{Te}}xt-I{\ul \emph{D}} semantic fusion approach for sequential {\ul \emph{Rec}}ommendation, namely \textbf{\our}. 
Different from existing side information fusion methods~\cite{hidasi2016gru4recf,zhang2019fdsa,gao2023smlp4rec}, our approach conducts the sequence-level semantic fusion on the transformed representations in the frequency domain. 
Specifically, \our~mainly consists of two key technical points.
(i) We inject the positional information by a mixture-of-experts~(MoE) enhanced adaptor, to improve  the discriminablity of text embeddings.  As such, it can learn more distinguishable textual representations for subsequent semantic fusion module.  
(ii) We propose a {mutual filter based method} to fuse ID and text embeddings in the frequency domain, and further employ an inverse FFT to produce {the integrated representations}. 
Such a transformation way has similar effects of contextual convolutions in the time domain and naturally integrates the bidirectional contextual information from the entire interaction sequence. 
Thus, it is more capable of capturing the sequential characteristics  and enhancing the text-ID semantic fusion. 
Our approach essentially provides a general framework to fuse multiple kinds of side information in recommender systems, which can be applied with various text encoders and recommender backbones.
Our contributions are summarized as follows:

$\bullet$ 
We propose a novel text-ID semantic fusion approach for sequential recommendation. The major contribution lies in that text embeddings and ID embeddings are transformed by Fourier Transform, and subsequently fused in the frequency domain. Such a fusion way can fully leverage sequential contexts and has the same effects of contextual convolution in the time domain.  

$\bullet$ 
We further propose to enhance the discriminability of the text embeddings from the text encoder, by adaptively injecting positional information via an MoE modulation method.   

$\bullet$ 
Extensive experiments are conducted on five public datasets, demonstrating the effectiveness of our proposed \our, notably with 14\% and 38\% performance gains over the competitive baselines on the ML-1M and OR datasets, respectively.
\section{Preliminary}
\label{sec:sec-preliminary}

\paratitle{Problem Statement.}
Let $\mathcal{U} = \{u_1, u_2, \ldots, u_{|\mathcal{U}|}\}$ denote a set of users and $\mathcal{V} = \{v_1, v_2, \ldots, v_{|\mathcal{V}|}\}$ denote a set of items.
Specifically, {each item $v$ is associated with a description text~(\eg product title),} 
and we denote it as $d_i = \{w_1, w_2, \ldots, w_c\}$, where each word $w$ is from some vocabulary and $c$ is the input length of the text.
Each user $u$ has an interaction context consisting of an ordered sequence of previously interacted items, \ie {$\{v_1, v_2, \ldots, v_n\}$}, where $n$ denotes the total number of interactions in the sequence. Given historical records, the sequential recommendation task can be defined as predicting the next item the user is likely to interact with at the ($n$ + 1)-th step, denoted as $p(v_{n+1}|\{v_1, v_2, \ldots, v_n\})$.

\paratitle{Fourier Transform.}
Discrete Fourier Transform~(DFT) is one of the classical methods in the field of sequence signal processing~\cite{rabiner1975theory}, which converts the sampled signal in the \emph{time domain} to the \emph{frequency domain}. 
Given the sequence data $\{x_j\}$ with $j \in \{1,\ldots,n-1\}$, DFT converts it into the frequency domain according to the formula:
\begin{equation}\label{eq:fft}
\mathcal{F}: \tilde{x}_k = \sum_{j=0}^{n-1} x_j\exp{\left(-\frac{2\pi i}{n}jk\right)}, 0 \leq k \leq n - 1,
\end{equation}
where $i$ is the imaginary unit and $\tilde{x}$ denotes the spectrum of the sequence $\{x_j\}$ at the frequency step $\omega_k={2\pi k}\big/{n}$.
Accordingly, given representations in the frequency domain, the following formula~(\emph{inverse DFT}) is used to recover signals to the time domain:
\begin{equation}\label{eq:ifft}
\mathcal{F}^{-1}: x_j = \frac{1}{n} \sum_{k=0}^{n-1} \tilde{x}_k\exp{\left(-\frac{2\pi i}{n}jk\right)}, 0 \leq j \leq n - 1.
\end{equation}

The Fast Fourier Transform~(FFT) is an efficient algorithm to calculate DFT, which is widely used in prior work~\cite{heideman1984gauss, van1992computational}. To improve the efficiency, FFT computes transformations by factorizing the DFT matrix into a product of sparse factors~\cite{van1992computational}, directly reducing the complexity from $\mathcal{O}(n^2)$ in DFT to $\mathcal{O}(n\log n)$ in FFT.

\section{Methodology}
\label{sec:sec-method}

\begin{figure*}[!t]
  \centering
  \captionsetup{font={small}}
  \includegraphics[width=.75\linewidth]{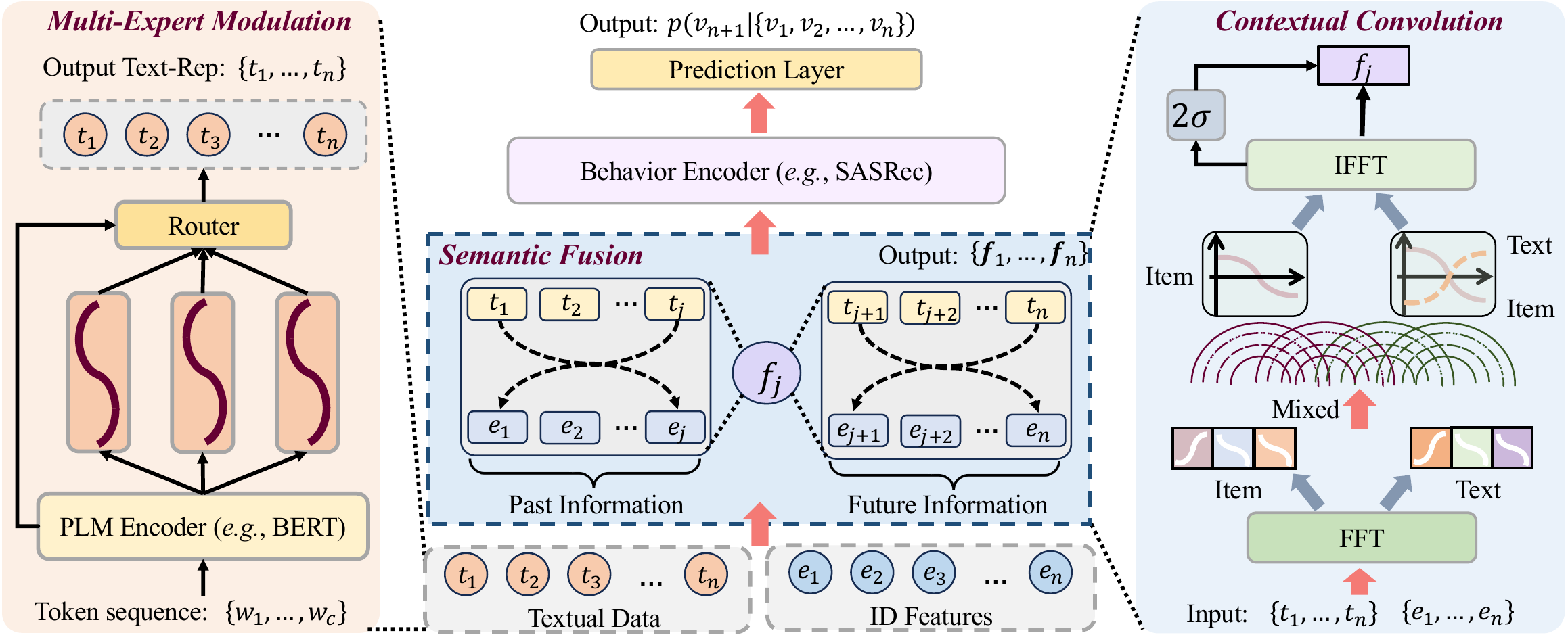}
  \caption{Overall architecture of our proposed \our.}
  \label{fig:model}
\end{figure*}

In this section, we present a novel {\ul \emph{Te}}xt-I{\ul \emph{D}} semantic fusion approach for sequential {\ul \emph{Rec}}ommendation~(as illustrated in Figure~\ref{fig:model}), namely \textbf{\our}. 
Next, we first introduce the detailed approach,  and then present theoretical analysis and discussion. 

\subsection{Sequence-Level Representation Fusion}
\label{sec:method-fusion}
To leverage textual information of items, previous work~\cite{hou2022unisrec,gao2023smlp4rec,liu2023dlfs-rec} typically integrates the text embedding from a text encoder and ID embedding from a recommender model in a relatively simple way, \eg concatenating the two embeddings at an item level.  
However, as discussed in Section~\ref{sec:intro}, such a way cannot effectively capture sequential contexts for semantic fusion, and the used text embeddings themselves lack discriminability among similar items.  
As our solution, we first propose a mixture-of-experts (MoE) enhanced modulation model, to learn more distinguishable textual representations (Section~\ref{sec:method-textual-repre}), and further develop a mutual filter based method to fuse the text and IDs in the frequency domain, to effectively capture the sequential contexts of semantic fusion (Section~\ref{sec:method-id-text-fusion}). 

\subsubsection{Distinguishable Textual Representation Encoding} 
\label{sec:method-textual-repre}
To obtain distinguishable text embeddings for recommendation, we employ a pre-trained language model~(PLM) for text encoding, and design a mixture-of-expert~(MoE) based modulation method to improve the discriminability of textual representations, {by incorporating sequential information with multiple sets of modulation embeddings}. 

\paratitle{PLM-based Text Encoding.} 
We utilize the widely used BERT~\cite{devlin2018bert} model to encode the textual data of items.
Following previous work~\cite{hou2022unisrec,hou2023vqrec,ding2021zesrec}, given the associated text of item $v$ with the length $c$ (\ie $\{w_1, w_2, \cdots, w_c\}$), we insert a $[\mathrm{CLS}]$ token at the beginning of the sequence,
and feed the extended sequence into the PLM: 
\begin{align}
    \bm t_v = \mathrm{PLM}(\{[\mathrm{CLS}; w_1; w_2; \cdots; w_c\}),
\end{align}
where $\bm t_v$ is the last hidden state vector of the input token ``$\mathrm{CLS}$'', and ``$[;]$'' denotes the concatenate operation.
In this way, each item $v$ is encoded with a unique \emph{text embedding} $\bm t_v$.
Note that the PLM is \emph{fixed}  during training, thus ensuring the efficiency of our approach.

\paratitle{Discriminability Enhancement.}
{Existing studies~\cite{hou2022unisrec,li2020sentence,huang2021whiteningbert}} have found that the text embeddings obtained by PLM (\eg BERT~\cite{devlin2018bert}) are less capable of discriminating the similar items. 
To address this issue, our solution is to employ a set of \emph{modulation embeddings} for enhancing the discriminability of textual representations. 
Formally, given a sequence of text embeddings for the items interacted by a user, \ie $\{\bm t_1, \cdots, \bm t_n\}$, the modulated representations are formulated as follows:
\begin{align}
    \bm t'_j &= (\bm t_j + \bm s_{j}) \cdot \bm W^P,
\end{align}
where  $\bm s_j$ is the modulation embedding of the $j$-th position, and $\bm W^P$ is the learnable parameter.
The employed modulation embeddings are similar to the absolute positional embeddings used in Transformer~\cite{vaswani2017attention}. 
In our approach, positional embeddings are intended for attention interaction based on the enhanced item representations \emph{after fusion}, while modulation embeddings are used to increase the discriminability of  textual representations \emph{before fusion}.  

\paratitle{Multi-Expert Modulation.}
Since interaction behaviors of users are very complex, the same textual representation might correspond to  varied sequential semantics~\cite{wang2021adaptive}. 
To effectively adapt to diverse interaction scenarios in recommender systems, we use the MoE architecture~\cite{hou2022unisrec} of multiple \emph{modulation experts} to further enhance the sequential discriminability of textual representations:  

\begin{align}\label{eq:moe}
    \bm t'_j &= \sum_{k = 1}^G g_k \cdot (\bm t_j + \bm s_{j, k}) \cdot \bm W^P_k,\\
    \bm g &= \mathrm{Softmax}(\bm t_j \cdot \bm W^G + \bm \delta),
\end{align}
where $\bm t'_j$ is the enhanced textual representation of the item in the $j$-th position, $\bm s_{j,k}$ is the modulation embedding for adjusting the textual frequency of the $j$-th position in the $k$-th expert, $G$ is the number of experts, $\bm W^P_k, \bm W^G$ are learnable parameters, $\bm g$ is the combination weight from the gating router and $\bm \delta$ is the random Gaussian noise for balancing the expert load.
Here, we expect that a mixture of modulation experts can capture diverse sequential contexts with adaptive positional information, thus leading to distinguishable textual representations for items in a sequence.   

\subsubsection{Text-ID Semantic Fusion in the Frequency Domain.}
\label{sec:method-id-text-fusion}
After obtaining the improved textual encoding $\bm t'$, we next discuss how to effectively fuse text and ID representations. 
Specially, we leverage the Fast Fourier Transform (see FFT in Eq.~\eqref{eq:fft}) to map the text and ID embeddings into the \emph{frequency domain}, enabling an effective integration of these two kinds of representations at a sequence level. 
Furthermore, we develop a gated fusion mechanism to adaptively  balance the importance of these two parts for semantic fusion. 
In what follows, we will introduce the details of each part.

\paratitle{Text-ID Mutual Filtering.}
In the semantic fusion layer, we fuse the text and ID embeddings in the frequency domain across each dimension. 
Formally, given the ID embeddings $\bm E = \{\bm e_1, \cdots, \bm e_n\} \in \mathbb{R}^{n \times d}$ and text embeddings $\bm T = \{\bm t_1', \cdots, \bm t_n'\} \in \mathbb{R}^{n \times d}$ (see Eq.~\eqref{eq:moe}), we first perform FFT (see Eq.~\eqref{eq:fft}) along each dimension to convert the $\bm T$ and $\bm E$ to the frequency domain as follows:

\begin{equation}
    \Tilde{\bm T} = \mathcal{F}(\bm T) \in \mathbb{C}^{n \times d}, \quad
    \Tilde{\bm E} = \mathcal{F}(\bm E) \in \mathbb{C}^{n \times d}, 
\end{equation}
where $\Tilde{\bm T}$ and $\Tilde{\bm E}$ denote the spectrum of text and ID embeddings, respectively. 
In this way, these two modalities can be mutually filtered by multiplying their spectrum in the frequency domain:
\begin{equation}\label{eq:element-multiply}
    \Tilde{\bm F} = \Tilde{\bm T} \odot \Tilde{\bm E} \in \mathbb{C}^{n \times d},
\end{equation}
where ``$\odot$'' denotes the element-wise multiplication, \ie Hadamard product.  
Furthermore, {we can also use a learnable filter $\bm W \in \mathbb{C}^{n \times d}$ to attenuate the noise of ID embeddings as follows: } 
\begin{equation}\label{eq:item-multiply}
    \Tilde{\bm E}' = \bm W \odot \Tilde{\bm E} \in \mathbb{C}^{n \times d},
\end{equation}
where $\Tilde{\bm F}$ and $\Tilde{\bm E}'$ are the fused representations and modulated ID features, respectively.
In the frequency domain, the global sequential characteristics of the original sequences can be inherently aggregated into the transformed representations (\ie $\Tilde{\bm E}$ and $\Tilde{\bm T}$). 
In this way, the multiplicative operations (\ie ``$\odot$'') can mimic the effect of convolution in the time domain (will be discussed in Section~\ref{sec:method-theoretical-analysis}), which essentially achieves the sequence-level semantic fusion. 
After mutual filtering, we adopt the inverse FFT (see Eq.~\eqref{eq:ifft}) to transform the representations $\Tilde{\bm E}'$ and $\Tilde{\bm F}$ to the original time domain: 
\begin{equation}
    {\bm F} \leftarrow \mathcal{F}^{-1}(\Tilde{\bm F}) \in \mathbb{R}^{n \times d}, \quad
    {\bm E}' \leftarrow \mathcal{F}^{-1}(\Tilde{\bm E}') \in \mathbb{R}^{n \times d}.
\end{equation} 
{With the above transformations, the outputs will ultimately consist of two parts of feature representations: 
one part is derived from the semantic fusion between text and ID embeddings (\ie ${\bm F}$), and the other part is derived from {the denoised ID embeddings} (\ie ${\bm E}'$).
} 
Finally, we develop a dual gating mechanism in the time domain to combine them in the following formula:
\begin{align}\label{eq:fused-rep}
    \bm V = 2 \sigma\Big(\mathrm{Gate}(\bm F)\Big) \cdot \bm F + 2 \sigma\Big(\mathrm{Gate}(\bm E')\Big) \cdot \bm E'.
\end{align}
Here $\mathrm{Gate}(\cdot): \mathbb{R}^{n \times d} \rightarrow \mathbb{R}$ is a linear function for generating the fusion weight, $\sigma$ is the sigmoid function, and a multiplier of 2 is used to transform the values to the range of $[0, 2]$ with an average of 1.
Hence, we use the final representations $\bm V = \{\bm v_1, \bm v_2, \cdots, \bm v_n\} \in \mathbb{R}^{n \times d}$ from Eq.~\eqref{eq:fused-rep} for subsequent user behavior modeling. 

\paratitle{Prediction and Optimization.}
Given a sequence of fused item representations (\ie $\bm V$ in Eq.~\eqref{eq:fused-rep}), we further utilize a user behavior encoder to obtain the sequence representation.
Note that our representation fusion approach can be integrated with various behavior encoders.
Taking the widely used Transformer architecture (\eg SASRec~\cite{kang2018sasrec}) as example, given the fused representations $\bm V = \{\bm v_1, \bm v_2, \cdots, \bm v_n\} \in \mathbb{R}^{n \times d}$ (see Eq.~\eqref{eq:fused-rep}), the sequential representation can be formulated as follows:
\begin{equation}\label{eq:attention}
    \bm x^0_j = \bm v_j + \bm p_j, \quad
    \bm X^{l+1} = \mathrm{FFN}\Big(\mathrm{MHAtt}(\bm X^l)\Big),
\end{equation}
where $\bm X^l = [\bm x_1^l, \bm x_2^l, \cdots, \bm x_n^l]$ is the output representation of the $l$-th layer, $\bm p_j$ is the absolute positional embedding of the $j$-th position, $\mathrm{FFN}(\cdot)$ is the point-wise feed-forward networks and $\mathrm{MHAtt}(\cdot)$ is the multi-head self-attention mechanism~\cite{vaswani2017attention}.
Given the representation of the last layer (\ie [$\hat{\bm x}_1, \hat{\bm x}_2, \cdots, \hat{\bm x}_n$]), we take the final hidden vector of the $n$-th position as the sequential representation (\ie $\hat{\bm x}_n$).
Finally, we adopt the widely used cross-entropy (CE) loss~\cite{sun2019bert4rec,hou2022unisrec} with a temperature parameter $\tau$ to train our model:
\begin{align}\label{eq:ce_loss}
    \mathcal{L}_{CE} = -\log\frac{\exp{(\bm \hat{\bm x}_n^{\top} \bm e_j / \tau)}}{\sum_{j' = 1}^{|\mathcal{V}|}\Big(\exp{(\bm \hat{\bm x}_n^{\top} \bm e_{j'} / \tau)\Big)}}.
\end{align}
Thus, we compute the probability of $v_{n+1}$ over the item set $|\mathcal{V}|$ as:
\begin{equation}\label{eq:predict_item}
    p\left(v_{n+1}|(v_1,v_2,…,v_n)\right) = \mathrm{Softmax}\left(\hat{\bm x}_n \cdot {\bm e}^{\top}\right),
\end{equation}

During training, the PLM encoder is fixed and 
we optimize the cross-entropy loss to train the backbone model, modulation experts as well as learnable filters via Eq.~\eqref{eq:ce_loss}. After training, we calculate the probability distribution of the softmax function for recommending the most possible item by Eq.~\eqref{eq:predict_item}.


\subsection{Theoretical Analysis}
\label{sec:method-theory}
As mentioned in Section~\ref{sec:intro}, previous work mainly suffers from two limitations in fusing representations of text and IDs \ie \emph{low discriminability} and \emph{item-level fusion}. 
In Section~\ref{sec:method-textual-repre}, we have discussed how to improve the discriminability of textual representations.
Furthermore, the fusion method proposed in Section~\ref{sec:method-id-text-fusion} can effectively integrate ID and textual representations at the sequence level. 
In this part, we theoretically demonstrate these superior properties of \our~ in \emph{sequence-level text-ID semantic fusion}~(Lemma~\ref{la:text-id-fusion}) and \emph{contextual user behavior integration}~(Lemma~\ref{la:expo}). 
\label{sec:method-theoretical-analysis}

\paratitle{Sequence-level Text-ID Semantic Fusion.} 
In the setting of our proposed \our~for sequential recommendation, the text-ID semantic fusion in the frequency domain can effectively integrate the contextual information of text and ID representations at the sequence level.
Formally, we can prove the following lemma to associate the frequency domain with the time domain:
\begin{lemma}\label{la:text-id-fusion}
The text-ID semantic fusion in the frequency domain (\ie $\mathcal{F}^{-1}_j(\Tilde{\bm T} \odot \Tilde{\bm E})$ in Eq.~\eqref{eq:element-multiply}) is equivalent to the text-ID contextual convolution in the time domain (\ie $\sum_{k = 0}^{n-1} {\bm t}_{k} \odot {\bm e}_{(j-k)\% n}$).
\end{lemma}
\paratitle{Proof}.  By the definition of Fourier transformation, we can obtain:
\begin{align}
\begin{split}    
    \bm f_j &= \mathcal{F}^{-1}_j\left(\Tilde{\bm t}_x \odot \Tilde{\bm e}_x\right)\\
    &= \mathcal{F}^{-1}_j\left(\sum_{k = 0}^{n-1} \exp\left(-\frac{2\pi i}{n} x k\right) {\bm t}_k \odot \sum_{m = 0}^{n-1} \exp\left(-\frac{2\pi i}{n} x m\right) {\bm e}_m\right)\\
    &= \mathcal{F}^{-1}_j\left(\sum_{k = 0}^{n-1}\sum_{m = 0}^{n-1} \exp\left(-\frac{2\pi i}{n} x (k + m)\right) {\bm t}_k \odot {\bm e}_m\right)\\
    &= \mathcal{F}^{-1}_j\left(\sum_{k = 0}^{n-1}\sum_{s = 0}^{n-1} \exp\left(-\frac{2\pi i}{n} x s\right) {\bm t}_{k} \odot {\bm e}_{(s-k)\% n}\right)\\   
    &= \mathcal{F}^{-1}_j\left(\sum_{s = 0}^{n-1}\sum_{k = 0}^{n-1} \exp\left(-\frac{2\pi i}{n} x s\right) {\bm t}_{k} \odot {\bm e}_{(s-k)\% n}\right)\\
    &= \mathcal{F}^{-1}_j\left(\mathcal{F}_x\left(\sum_{k = 0}^{n-1} {\bm t}_{k} \odot {\bm e}_{(s-k)\% n}\right)\right) = \sum_{k = 0}^{n-1} {\bm t}_{k} \odot {\bm e}_{(j-k)\% n},
\end{split}\label{eq:pro1}
\end{align}
where ``$\%$'' is the modular operation, $\bm f_j$ is the \emph{circular convolution} operation, which aggregates the pair-wise similarity between the textual data $\bm t_k$ and the ID feature $\bm e_{j - k}$ within the given sequence. 
Similar to CNN~\cite{krizhevsky2012imagenet-cnn,tang2018caser},  our proposed \emph{mutual filtering operation by dimension} (Eq.~\eqref{eq:element-multiply}) is essentially a \emph{convolution operation over the sequences} of ID and text embeddings. 
Therefore, it can effectively capture the global contexts of the entire interaction sequence for  semantic fusion (before attention interaction in Transformer). 

\paratitle{Contextual Integration.} 
As proved in Lemma~\ref{la:text-id-fusion},  the proposed approach can achieve sequence-level semantic fusion. In this part, we further study the specific forms of contextual integration possessed by our approach. 
Formally, we use the function $S(m, n) = \sum_{j = m}^{n} = {\bm t}_{j} \odot {\bm e}_{(m + n - j)\% n}$ to describe the sequential information between the $m$-th position and $n$-th position, \eg $\mathcal{S}(1,2) = \bm t_1 \odot \bm e_2 + \bm t_2 \odot \bm e_1, \mathcal{S}(n,n) = \bm t_n \odot \bm e_n$.
We have the following observation:

\begin{lemma}\label{la:expo}
Given the ID embeddings $\{\bm e_1, \cdots, \bm e_n\}$ and the text embeddings $\{\bm t_1, \cdots, \bm t_n\}$,  the fused representations can be denoted as $\{\bm f_1, \cdots, \bm f_n\}$, where each element can be modeled by combining both the past and future information of the sequence, \ie $\bm f_j = \mathcal{S}(0, j) + \mathcal{S}(j + 1, n - 1)$.
\end{lemma}
\paratitle{Proof.}  According to Eq.~\eqref{eq:pro1}, for the $j$-th position, we obtain:
\begin{align}
\begin{split}  
        \bm f_j & = \sum_{k = 0}^{n-1} {\bm t}_{k} \odot {\bm e}_{(j-k)\% n},\\
         & = \sum_{k = 0}^j {\bm t}_{k} \odot {\bm e}_{(j-k)\% n} + \sum_{k = j + 1}^{n-1} {\bm t}_{k} \odot {\bm e}_{(j-k)\% n},\\
         & = \sum_{k = 0}^j {\bm t}_{k} \odot {\bm e}_{(j-k)} + \sum_{k = j + 1}^{n-1} {\bm t}_{k} \odot {\bm e}_{(j + n -k)},\\
         & =  \underbrace{\mathcal{S}(0, j)}_{\text{Past}} + \underbrace{\mathcal{S}(j + 1, n - 1)}_{\text{Future}}.
\end{split}\label{eq:pro2}
\end{align}
It shows that our proposed convolution operation at some specific position is capable of integrating both the \emph{past} and \emph{future} sequential information, which is similar to the bidirectional sequential modeling in BERT~\cite{devlin2018bert}. 
Such a property is particularly important for modeling sequential user behavior data,  which may also  possess some potential merits. 
For example, since $\mathcal{S}$ is only related to the sequential context, the sequences with the same preceding or subsequent parts will share partial representations, which can be utilized to improve the generalization capabilities of predictive models (\eg reusing the learned partial representations for a new sequence).  
These properties will be investigated in future work.


\subsection{Discussion}
\label{sec:method-discussion}


\begin{table}[t]
\small
\centering
\captionsetup{font={small}}
\caption{Comparison of different methods. ``Distinguishable'' denotes that textual representations are enhanced for alignment and uniformity. ``Sequence-level'' fusion considers cross-item fusion within a sequence, and ``MA'' denotes the model-agnostic fusion.}
\label{tab:method-compare}
\begin{tabular}{@{}ccccc@{}}
\toprule
Method   & Fusion Type                     & Disting. & Seq.-level & MA \\ \midrule
SASRecF~\cite{hidasi2016gru4recf}  & Representation concat.    & \color[RGB]{224,30,55}{\XSolidBrush}       & \color[RGB]{224,30,55}{\XSolidBrush}            & \color[RGB]{5,130,202}{\CheckmarkBold}            \\
FDSA~\cite{zhang2019fdsa}     & Logits concat.            & \color[RGB]{224,30,55}{\XSolidBrush}       & \color[RGB]{224,30,55}{\XSolidBrush}            & \color[RGB]{5,130,202}{\CheckmarkBold}            \\
S$^3$Rec~\cite{zhou2020s3rec} & Attribute pre-training          & \color[RGB]{224,30,55}{\XSolidBrush}       & \color[RGB]{5,130,202}{\CheckmarkBold}            & \color[RGB]{224,30,55}{\XSolidBrush}            \\
DIF-SR~\cite{xie2022dif-sr}   & Decomposed attention            & \color[RGB]{224,30,55}{\XSolidBrush}       & \color[RGB]{224,30,55}{\XSolidBrush}            & \color[RGB]{224,30,55}{\XSolidBrush}            \\
UniSRec~\cite{hou2022unisrec}  & Vector addition                 & \color[RGB]{5,130,202}{\CheckmarkBold}       & \color[RGB]{224,30,55}{\XSolidBrush}            & \color[RGB]{5,130,202}{\CheckmarkBold}            \\
DLFS-Rec~\cite{liu2023dlfs-rec} & Distribution concat.      & \color[RGB]{224,30,55}{\XSolidBrush}       & \color[RGB]{224,30,55}{\XSolidBrush}            & \color[RGB]{224,30,55}{\XSolidBrush}            \\
SMLP4Rec~\cite{gao2023smlp4rec} & Representation concat.    & \color[RGB]{224,30,55}{\XSolidBrush}       & \color[RGB]{5,130,202}{\CheckmarkBold}            & \color[RGB]{224,30,55}{\XSolidBrush}            \\
\our     & Contextual convolution & \color[RGB]{5,130,202}{\CheckmarkBold}       & \color[RGB]{5,130,202}{\CheckmarkBold}            & \color[RGB]{5,130,202}{\CheckmarkBold}            \\ \bottomrule
\end{tabular}
\end{table}

\paratitle{Comparison with Existing Work}. As shown in Table~\ref{tab:method-compare}, we compare several sequential models with side information fusion to highlight our novelty and differences. 
First, previous studies on side information fusion typically perform item-level fusion such as representation concatenation~\cite{hidasi2016gru4recf,gao2023smlp4rec}, while our work conducts text-ID semantic fusion in the frequency domain through Fourier transformation. As proved in Lemma~\ref{la:text-id-fusion} and~\ref{la:expo}, the proposed approach can effectively 
fuse the semantics of ID and text embeddings at a sequence level, which has the similar effect of sequential convolution.  
Secondly, we propose to employ MoE enhanced modulation approach to enhance the discriminability of the text embeddings from the PLM encoder, which injects positional information to text embeddings via multiple sets of modulation embeddings.    
Thirdly, our approach is general and flexible to work with various text encoders and recommender backbones, which has an extensive applicability in existing recommender systems (see Section~\ref{sec:exp-further-analysis}).  
\ignore{First, UniSRec~\cite{hou2022unisrec} and our method \textcolor{blue}{can utilize the semantic associated data with heterogeneous textual representations}. Secondly, besides the time domain for sequential patterns, DLFS-Rec and~\our~cast representation of items into the frequency domain with learnable filters, \textcolor{blue}{enabling representations to be fully integrated}. Thirdly, our sequence-level fusion strategy considers the global representation fusion across ID and text features within the entire sequence, which cannot be achieved by the locally item-level fusion, \eg vector addition in UniSRec.
\textcolor{blue}{Finally, our model-agnostic framework can be adapted to different model backbones, while DIF-SR~\cite{xie2022dif-sr}, DLFS-Rec and SMLP4Rec~\cite{gao2023smlp4rec} are tailored to specific models.} Further, our proposed fusion module is essentially independent of the used text encoder (\eg BERT). \textcolor{blue}{These designs enable our model to be seamlessly integrated with mainstream algorithms with a wide choice of textual representation methods}~(see Section~\ref{sec:exp-further-analysis}).}


\paratitle{Complexity Analysis}. Our model mainly utilizes operations of FFT in Eq.~\eqref{eq:fft} and inverse FFT in Eq.~\eqref{eq:ifft} to conduct the space transformation, with the time complexity of $\mathcal{O}(nd\log n)$ as discussed in Section~\ref{sec:sec-preliminary}, where $d$ is the hidden dimension and $n$ is the number of interacted items within an interaction sequence.
In Section~\ref{sec:method-textual-repre}, the cost of frequency modulation is $\mathcal{O}(Gnd)$ and $G$ denotes the number of experts.
As for text-ID semantic fusion in Section~\ref{sec:method-id-text-fusion}, the time complexity of element-wise multiplication in Eq.~\eqref{eq:element-multiply} and gating mechanism in Eq.~\eqref{eq:fused-rep} is $\mathcal{O}(nd)$.
Therefore, the complexity of textual representation fusion in~\our~is $\mathcal{O}(n(G+\log n)d)$. In general, our proposed method is comparable to existing approaches~\cite{hou2022unisrec,gao2023smlp4rec} since $d$, $G$ and $n$ are constants independent of the model architecture.
In terms of the model backbone~(\eg SASRec), our plug-and-play framework has no additional time cost to the self-attention mechanism for modeling sequential patterns in Eq.~\eqref{eq:attention}.
Furthermore, our approach does not require any additional training of PLMs, which ensures the efficiency of our proposed~\our.

\section{Experiment}

\begin{table}[t]
\small
\centering
\captionsetup{font={small}}
\caption{Statistics of the processed datasets.}
\label{tab:exp-datasets}
\begin{tabular}{@{}crrrc@{}}
\toprule
\textbf{Datasets} & \multicolumn{1}{c}{\textbf{\#Users}} & \multicolumn{1}{c}{\textbf{\#Items}} & \multicolumn{1}{c}{\textbf{\#Interactions}} & \textbf{Sparsity} \\ \midrule
ML-1M             & 6,040                                & 3,416                                & 993,571                                     & 95.1867\%         \\
OR                & 16,520                               & 3,469                                & 503,386                                     & 99.1219\%         \\
Office            & 87,346                               & 25,986                               & 597,491                                     & 99.9737\%         \\
Food              & 115,349                              & 39,670                               & 912,064                                     & 99.9801\%         \\
Movies            & 281,700                              & 59,203                               & 2,945,031                                   & 99.9823\%         \\ \bottomrule
\end{tabular}
\end{table}


\subsection{Experimental Setting} 
\label{sec:exp-settings}

\subsubsection{Dataset Descriptions.} As shown in Table~\ref{tab:exp-datasets}, our experiments are conducted on five public benchmark datasets varying in platform, scale and sparsity. \textbf{MovieLens-1M}~\cite{harper2015movielens} is one of the most widely used datasets in recommender systems, which collects one million ratings of movies from the MovieLens website. 
We concatenate the fields of \emph{title}, \emph{genre} and \emph{year} of a movie as the descriptive text. \textbf{Online Retail}~(OR)~\cite{chen2012online-retail} is a transnational dataset containing transaction records from an e-commerce platform in UK, and the \emph{description} field is utilized as the text of items.
\textbf{Office}, \textbf{Food} and \textbf{Movies} are three subsets from the real-world product reviews in Amazon~\cite{ni2019amazon}, and we adopt the up-to-date version released in 2018. Following previous studies~\cite{hou2022unisrec,hou2023vqrec}, we concatenate fields of \emph{title}, \emph{categories} and \emph{brand} as the textual feature. In line with the existing literature~\cite{kang2018sasrec,zhou2020s3rec,liu2023dlfs-rec}, we apply the five-core strategy to filter inactive users and unpopular items with fewer than five records. 


\begin{table*}[t]
\small
\centering
\captionsetup{font={small}}
\caption{Overall Performance Comparison. The best and runner-up results are bold and underlined, respectively. ``Impr'' means the improvement of our proposed~\our~over the best baseline, and ``*'' denotes that improvements are significant at the level of 0.01 with paired t-test. ``R@K'' and ``N@K'' stand for Recall@K and NDCG@K, respectively. }
\label{tab:exp-over-performance}
\begin{tabular}{@{}cccccccccccccr@{}}
\toprule
Dataset                 & Metric & SASRec & FMLP   & FEARec       & SASRecF      & FDSA   & S$^3$Rec  & DIF-SR & UniSRec      & DLFS-Rec         & SMLP4Rec         & Ours             & Impr.    \\ \midrule
\multirow{4}{*}{ML-1M}  & R@10   & 0.2300 & 0.2363 & {\ul 0.2407} & 0.2356       & 0.2286 & 0.2268 & 0.2313 & 0.2257       & 0.2179       & 0.2157       & \textbf{0.2623*} & +8.97\%  \\
                        & R@20   & 0.3439 & 0.3500 & {\ul 0.3549} & 0.3515       & 0.3435 & 0.3431 & 0.3507 & 0.3474       & 0.3068       & 0.3346       & \textbf{0.3709*} & +4.45\%  \\
                        & N@10   & 0.1146 & 0.1234 & {\ul 0.1258} & 0.1249       & 0.1170 & 0.1138 & 0.1164 & 0.1140       & 0.1238       & 0.1107       & \textbf{0.1445*} & +14.86\% \\
                        & N@20   & 0.1433 & 0.1520 & {\ul 0.1554} & 0.1541       & 0.1460 & 0.1430 & 0.1464 & 0.1446       & 0.1460       & 0.1407       & \textbf{0.1719*} & +10.62\% \\ \midrule
\multirow{4}{*}{OR}     & R@10   & 0.1545 & 0.1492 & 0.1532       & 0.1479       & 0.1491 & 0.1464 & 0.1522 & 0.1526       & {\ul 0.1634} & 0.1515       & \textbf{0.2234*} & +36.72\% \\
                        & R@20   & 0.2496 & 0.2490 & {\ul 0.2509}       & 0.2344       & 0.2369 & 0.2395 & 0.2478 & 0.2421       & 0.2285       & {0.2508} & \textbf{0.3073*} & +22.48\% \\
                        & N@10   & 0.0699 & 0.0690 & 0.0775       & 0.0744       & 0.0709 & 0.0692 & 0.0708 & 0.0717       & {\ul 0.0952} & 0.0650       & \textbf{0.1316*} & +38.24\% \\
                        & N@20   & 0.0939 & 0.0923 & 0.0971       & 0.0962       & 0.0931 & 0.0951 & 0.0948 & 0.0943       & {\ul 0.1115} & 0.0901       & \textbf{0.1527*} & +36.95\% \\ \midrule
\multirow{4}{*}{Office} & R@10   & 0.1061 & 0.1138 & 0.1172       & 0.1081       & 0.1111 & 0.1111 & 0.1162 & {\ul 0.1233}       & 0.1131       & {0.1183} & \textbf{0.1356*} & +9.98\% \\
                        & R@20   & 0.1272 & 0.1367 & 0.1418       & 0.1276       & 0.1329 & 0.1322 & 0.1406 & {\ul 0.1510}       & 0.1291       & {0.1483} & \textbf{0.1598*} & +5.83\%  \\
                        & N@10   & 0.0699 & 0.0723 & 0.0757       & 0.0841       & 0.0851 & 0.0764 & 0.0783 & 0.0723       & {\ul 0.0916} & 0.0700       & \textbf{0.1052*} & +14.85\% \\
                        & N@20   & 0.0753 & 0.0781 & 0.0792       & 0.0890       & 0.0906 & 0.0818 & 0.0844 & 0.0792       & {\ul 0.0957} & 0.0776       & \textbf{0.1113*} & +16.30\% \\ \midrule
\multirow{4}{*}{Food}   & R@10   & 0.1069 & 0.1133 & 0.1192       & 0.1084       & 0.1100 & 0.1074 & 0.1144 & {\ul 0.1259} & 0.1146       & 0.1160       & \textbf{0.1327*} & +5.40\%  \\
                        & R@20   & 0.1281 & 0.1385 & 0.1395       & 0.1285       & 0.1311 & 0.1283 & 0.1375 & {\ul 0.1543} & 0.1329       & 0.1475       & \textbf{0.1604*} & +3.95\%  \\
                        & N@10   & 0.0736 & 0.0745 & 0.0780       & 0.0841       & 0.0846 & 0.0739 & 0.0786 & 0.0733       & {\ul 0.0912} & 0.0688       & \textbf{0.1012*} & +10.96\% \\
                        & N@20   & 0.0789 & 0.0809 & 0.0835       & 0.0892       & 0.0899 & 0.0792 & 0.0844 & 0.0805       & {\ul 0.0958} & 0.0768       & \textbf{0.1082*} & +12.94\% \\ \midrule
\multirow{4}{*}{Movies} & R@10   & 0.1453 & 0.1477 & 0.1496       & 0.1425       & 0.1446 & 0.1461 & 0.1440 & 0.1493       & {\ul 0.1504} & 0.1425       & \textbf{0.1611*} & +7.11\%  \\
                        & R@20   & 0.1831 & 0.1894 & 0.1897       & 0.1775       & 0.1818 & 0.1822 & 0.1821 & {\ul 0.1907} & 0.1883       & 0.1865       & \textbf{0.1998*} & +4.77\%  \\
                        & N@10   & 0.0962 & 0.0975 & 0.0966       & {\ul 0.1040} & 0.1028 & 0.0964 & 0.0972 & 0.0858       & 0.1038       & 0.0921       & \textbf{0.1188*} & +14.23\% \\
                        & N@20   & 0.1058 & 0.1061 & 0.1067       & {\ul 0.1129} & 0.1121 & 0.1065 & 0.1068 & 0.0963       & 0.1118       & 0.1032       & \textbf{0.1285*} & +13.82\% \\ \bottomrule
\end{tabular}
\end{table*}

\paratitle{Baselines.}
We compare \our~with several approaches as follows.

\begin{itemize}[leftmargin=*]
    \item \textbf{SASRec}~\cite{kang2018sasrec} is the first sequential recommender based on the unidirectional self-attention mechanism. 
    \item \textbf{FMLP-Rec}~\cite{zhou2022fmlp-rec} introduces learnable filters to replace the self-attention mechanism with vanilla multilayer perceptron~(MLP).
    \item \textbf{FEARec}~\cite{du2023fearec} proposes the frequency enhanced attention network, and contrastive learning is utilized to align representations.
    \item \textbf{SASRecF}~\cite{hidasi2016gru4recf} concatenates representations of both items and item attributes as the fused input to extend SASRec.
    \item \textbf{FDSA}~\cite{zhang2019fdsa} utilizes two different Transformer encoders to
    encode items and features, respectively. Then it concatenates attention outputs from two encoders as the final output by late fusion.
    \item \textbf{S$^3$Rec}~\cite{zhou2020s3rec} is the first work to incorporate self-supervised learning in sequential recommendation with four pre-training tasks.
    \item \textbf{DIF-SR}~\cite{xie2022dif-sr} improves the side information fusion in Transformer-based recommendation by decoupled self-attention mechanism.
    \item \textbf{UniSRec}~\cite{hou2022unisrec} utilizes the textual descriptions of items to learn transferable representations. For a fair comparison, we train UniSRec from scratch without pre-training on additional datasets. 
    \item \textbf{DLFS-Rec}~\cite{liu2023dlfs-rec} regards representations of items and side information as the Gaussian distribution, and proposes distribution-based learnable filters for modeling sequences of items.
    \item \textbf{SMLP4Rec}~\cite{gao2023smlp4rec} captures sequential, cross-channel, and cross-feature dependencies for tri-directional side information fusion.
\end{itemize}



\subsubsection{Evaluation Metrics.} To evaluate the performance of next-item prediction, we apply the leave-one-out strategy~\cite{sun2019bert4rec,zhou2020s3rec,gao2023smlp4rec} to split the interaction sequence of one user into training, validation and test sets, respectively. We compute metrics based on the full-sort protocol that ranks the ground-truth item among all items, and results are averaged over all users~\cite{hou2022unisrec,du2023fearec}. For evaluation metrics, we utilize two commonly adopted ranking metrics Recall@K and NDCG@K~\cite{NDCG}, where K is set to 10 and 20 in our experiments. 

\subsubsection{Implementation Details.} 
We implement all models based on the open-source benchmark library \textsc{RecBole}~\cite{recbole[1.0]}. 
To ensure a fair comparison, we optimize models with the Adam optimizer and cross-entropy loss, 
and all experiments are conducted on the NVIDIA A100 machine. 
For all models, the training batch size is set to 2,048, and we adopt the early-stopping strategy to finish training when NDCG@10 on the validation set does not improve for 10 epochs.
We also tune the learning rate in \{5e-4,1e-3,5e-3\} for the optimal performance. 
For baselines, we carefully search the hyper-parameters following original papers. 
Furthermore, we provide our code, datasets and logged results to improve reproducibility at this repository: \textcolor{magenta}{\url{https://github.com/RUCAIBox/TedRec}}. 

\subsection{Performance Comparison}
\label{sec:exp-performance-compare}

As shown in Table~\ref{tab:exp-over-performance}, we compare~\our~with baselines on five public datasets, and empirical findings are summarized as follows.

For the ID-based models, SASRec~\cite{kang2018sasrec} and FEARec~\cite{du2023fearec} utilize the self-attention module, while FMLP-Rec~\cite{zhou2022fmlp-rec} and FEARec consider frequency-enhanced representations. 
The performance of MLP-based FMLP-Rec is better than that of SASRec, demonstrating the superiority of filtering algorithms~(\eg Band-Stop Filter) on attenuating the noise information of user interactions~\cite{du2023fearec,zhou2022fmlp-rec,liu2023dlfs-rec}. Furthermore, FEARec performs better than FMLP-Rec due to the attention mechanism in the frequency domain and contrastive learning techniques.
Meanwhile, FEARec based on the pure IDs is comparable to text-enhanced models on the dense MovieLens-1M dataset, indicating the importance of fully modeling textual features. 

For sequential models with textual information fusion, SASRecF~\cite{hidasi2016gru4recf}, FDSA~\cite{zhang2019fdsa}, S$^3$Rec~\cite{zhou2020s3rec} and DIF-SR~\cite{xie2022dif-sr} integrate the item texts into the self-attention framework from perspectives of early fusion, late fusion, attribute pre-training, and decoupled attention, respectively. These four models have varying degrees of improvement compared to SASRec on different datasets, showcasing the effect of 
auxiliary item information in sequential recommendation. 
Further, our method outperforms UniSRec by a large margin since we propose the multi-expert modulation and contextual convolution modules to effectively fuse IDs and texts.
As for SMLP4Rec~\cite{gao2023smlp4rec}, this approach fuses different views of sequences in the time domain, but fails to capture fine-grained periodicity due to the existence of noise. 
DLFS-Rec~\cite{liu2023dlfs-rec} is the most competitive baseline, because it adopts Gaussian distributions to represent items and additional features, and models item sequences with learnable filters. However, it still uses item-level operations such as concatenation to combine two kinds of distributions, limiting the flexibility of sequence-level fusion for frequency enhanced representations.
In general, the superior performance of our method against all baselines on five datasets verifies the effectiveness of~\our.

\subsection{Further Analysis}
\label{sec:exp-further-analysis}


\subsubsection{Ablation Study} In this part, we examine whether each of our proposed components plays a positive effect in the final performance. As for the ablation study, we analyze the following four variants of our method for comparison. (1) {\ul \emph{w/o} MM} replaces the multi-expert modulation in our textual representations~(Eq.~\eqref{eq:moe}) with the item encoding architecture in UniSRec, which adopts MoE-enhanced parametric whitening without our modulation embedding. (2) {\ul \emph{w/o} AG} replaces the adaptive gate in contextual convolution with direct additions in Eq.~\eqref{eq:fused-rep}. (3) {\ul \emph{w/o} IF} removes the item convolution in the frequency domain with pure item embeddings. (4) {\ul \emph{w/o} TF} removes the contextual convolution in the frequency domain with the vector product of text and IDs in the time domain.

\begin{figure}[t]
\centering
    \begin{minipage}[t]{0.23\textwidth}
        \centering
        \includegraphics[width=1\textwidth]{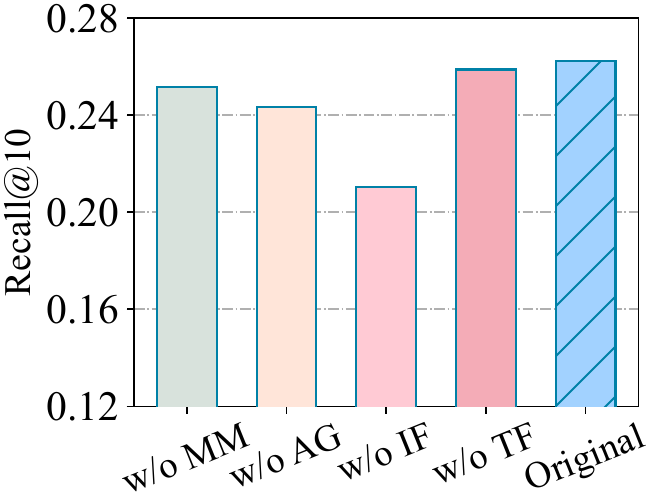}
        \captionsetup{font={small}}
        \subcaption{MovieLens-1M}
        \label{fig:exp-ablation-ml}
    \end{minipage}
    \begin{minipage}[t]{0.23\textwidth}
        \centering
        \includegraphics[width=1\textwidth]{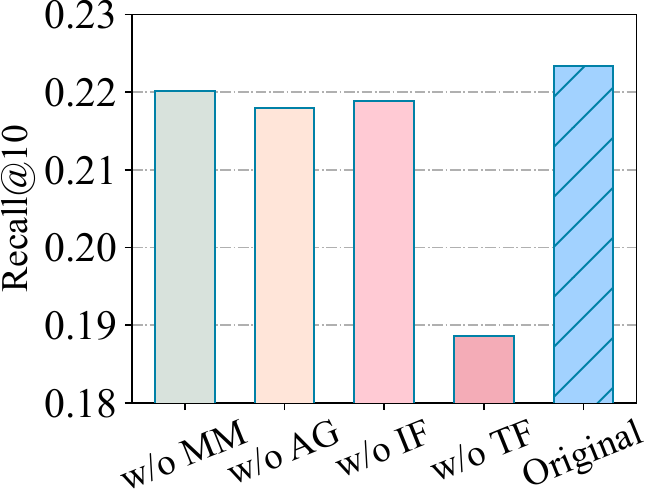}
        \captionsetup{font={small}}
        \subcaption{Online Retail} 
        \label{fig:exp-ablation-or}
        \end{minipage}
    \captionsetup{font={small}}
    \caption{Ablation study of our variants.} 
    \label{fig:exp-ablation}
\end{figure}

The performance comparison of our method with four variants is illustrated in Figure~\ref{fig:exp-ablation}. It can be observed that all the proposed components in~\our~affect the overall recommendation performance, and our final framework achieves the best. Moreover, on the sparse Online Retail dataset, the variant {\ul \emph{w/o} TF} gets poor results since the sequence-level representation fusion plays an important role in providing semantic attributes for collaborative information.


\subsubsection{Efficiency Comparison}

\begin{figure}[t]
\centering
    \begin{minipage}[t]{0.23\textwidth}
        \centering
        \includegraphics[width=1\textwidth]{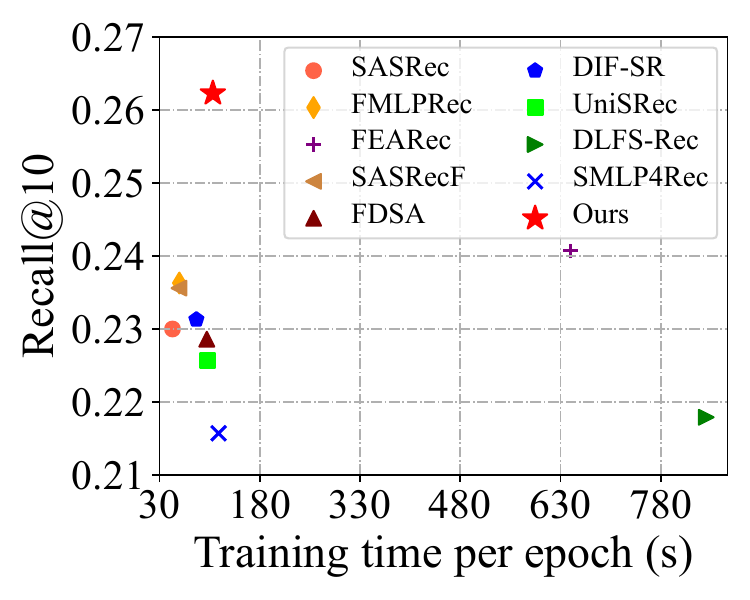}
        \captionsetup{font={small}}
        \subcaption{MovieLens-1M}
        \label{fig:exp-efficiency-ml}
    \end{minipage}
    \begin{minipage}[t]{0.23\textwidth}
        \centering
        \includegraphics[width=1\textwidth]{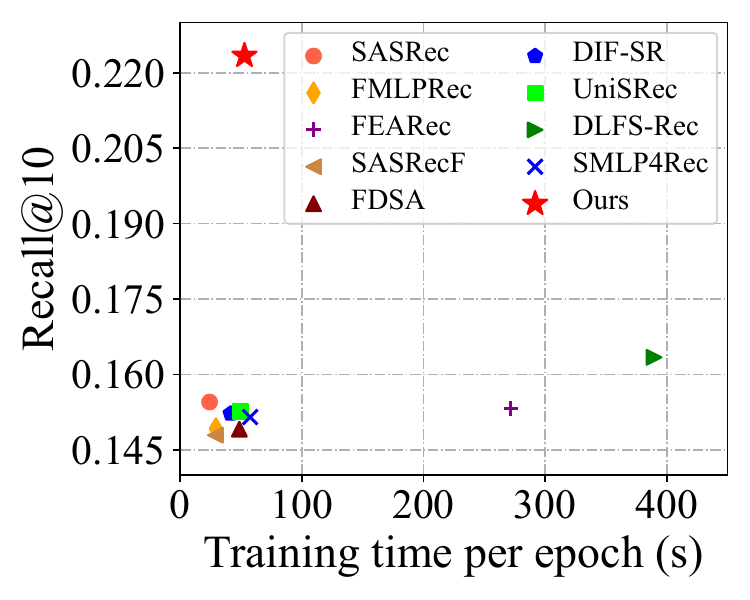}
        \captionsetup{font={small}}
        \subcaption{Online Retail} 
        \label{fig:exp-efficiency-or}
        \end{minipage}
    \captionsetup{font={small}}
    \caption{Performance comparison \wrt training efficiency.} 
    \label{fig:exp-efficiency}
\end{figure}

To further analyze the training efficiency of our proposed method, we plot scatter charts of efficiency~(training time per epoch) and effectiveness~(Recall@10) on two datasets. As shown in Figure~\ref{fig:exp-efficiency}, we can observe that our method has comparable efficiency with baseline models, while there is a significant improvement \wrt recommendation performance. In terms of the competitive baselines FEARec~\cite{du2023fearec} and DLFS-Rec~\cite{liu2023dlfs-rec}, despite their outstanding recommendation performance compared to other baselines, FEARec and DLFS-Rec have the common problem of long training time. The reason is that reconstruction of training data in contrastive learning~(\ie FEARec) and Wasserstein distance calculation in Gaussian distributions~(\ie DLFS-Rec) are time-consuming, limiting the training efficiency of practical applications.

\subsubsection{Impact of User Groups} 

\begin{figure}[t]
\centering
    \begin{minipage}[t]{0.23\textwidth}
        \centering
        \includegraphics[width=1\textwidth]{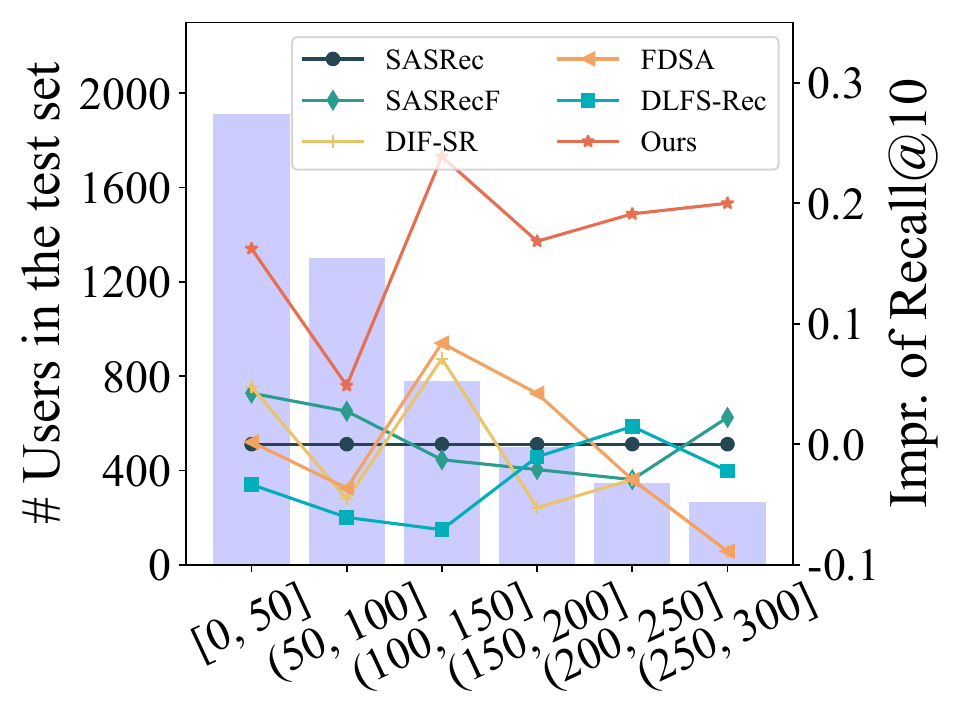}
        \captionsetup{font={small}}
        \subcaption{MovieLens-1M}
        \label{fig:exp-group-ml}
    \end{minipage}
    \begin{minipage}[t]{0.23\textwidth}
        \centering
        \includegraphics[width=1\textwidth]{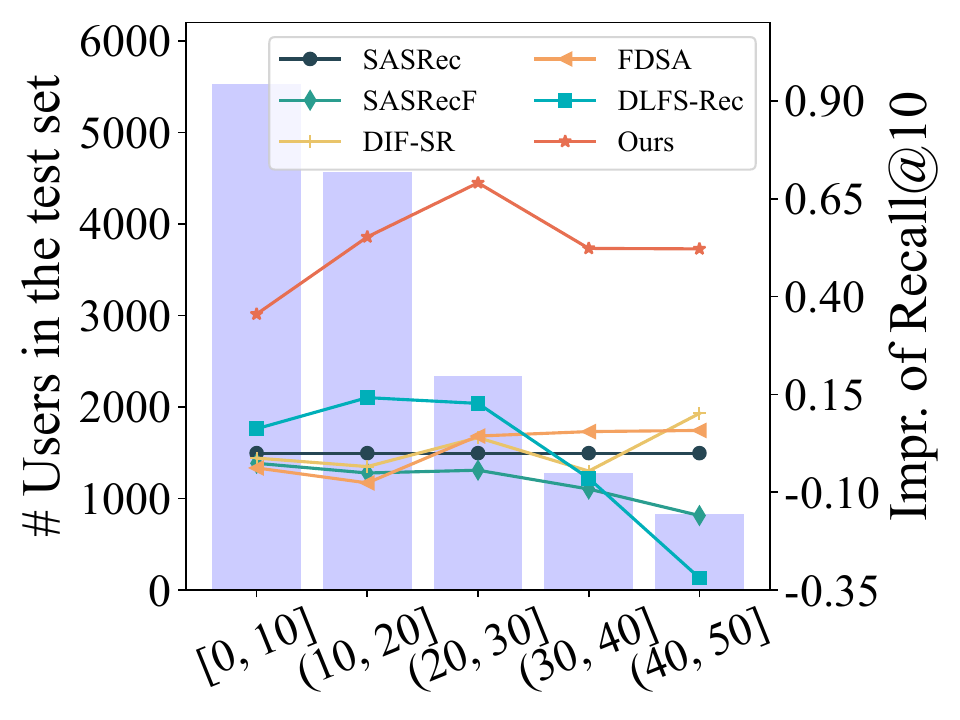}
        \captionsetup{font={small}}
        \subcaption{Online Retail} 
        \label{fig:exp-group-or}
        \end{minipage}
    \captionsetup{font={small}}
    \caption{Performance comparison for users with different levels of sparsity. The line chart denotes the improvements of corresponding models compared to SASRec \wrt Recall@10, while the bar chart represents the number of users in the test set considering interactions within a specified interval.} 
    \label{fig:exp-group}
\end{figure}

To analyze the impact of our method on user groups with different user activity, we divide all users in the test set into different groups based on their interactions in the training set. As shown in Figure~\ref{fig:exp-group}, we can observe that the improvements of our method compared to baselines with textual features are significant among all groups on two datasets. Especially, the bar chart indicates that both two datasets present a long-tail distribution. For inactive users with fewer interactions, our proposed~\our~still achieves performance improvement compared to other baselines, indicating the superiority of our fusion scheme.

\subsubsection{Impact of Model Backbones} 

To verify the model-agnostic property and generalization ability of our method, we implement our framework on four model backbones and report the performance difference. As shown in Table~\ref{tab:exp-model-backbone}, we can observe that our proposed method can consistently improve the performance of UniSRec, DLFS-Rec, SASRecF and DIF-SR, validating the effectiveness of~\our~on different models. 
Furthermore, the improvements on UniSRec and DLFS-Rec are significant compared to SASRecF and DIF-SR. A possible reason is that SASRecF and DIF-SR do not consider textual modeling or expert modulation, which limits the effectiveness of our frequency-aware semantic fusion method. 

\begin{table}[t]
\small
\centering
\captionsetup{font={small}}
\caption{Performance comparison \wrt different models. }
\label{tab:exp-model-backbone}
\begin{tabular}{@{}ccccc@{}}
\toprule
\multirow{2}{*}{Method} & \multicolumn{2}{c}{MovieLens-1M}  & \multicolumn{2}{c}{Online Retail} \\ \cmidrule(l){2-5} 
                        & Recall@10       & NDCG@10         & Recall@10       & NDCG@10         \\ \midrule
UniSRec                 & 0.2257          & 0.1140          & 0.1526          & 0.0717          \\
\emph{w/} ours          & \textbf{0.2623} & \textbf{0.1445} & \textbf{0.2234} & \textbf{0.1316} \\ \midrule
DLFS-Rec                & 0.2179          & 0.1238          & 0.1634          & 0.0952          \\
\emph{w/} ours          & \textbf{0.2513} & \textbf{0.1386} & \textbf{0.2248} & \textbf{0.1313} \\ \midrule
SASRecF                 & 0.2391          & 0.1267          & 0.1446          & 0.0700          \\
\emph{w/} ours          & \textbf{0.2427} & \textbf{0.1292} & \textbf{0.1666} & \textbf{0.0825} \\ \midrule
DIF-SR                  & 0.2313          & 0.1164          & 0.1522          & 0.0708          \\
\emph{w/} ours          & \textbf{0.2354} & \textbf{0.1284} & \textbf{0.1666} & \textbf{0.0893} \\ \bottomrule
\end{tabular}
\end{table}

\subsubsection{Impact of Textual Representations} 

As shown in Table~\ref{tab:lm-text-repre}, we compare the performance of four language models on \our, namely BERT~\cite{devlin2018bert}, T5~\cite{raffel2020T5}, Flan-T5~\cite{chung2022flan-t5}, and LLaMA2~\cite{touvron2023llama2}. Besides, we use random vectors as textual representations~(denoted by ``Random'') to highlight the semantic merits of language models.
In general, the ``CLS'' embedding from pre-trained BERT achieves the best performance among four models, and the possible reason is that the mean pooling of tokens in a text sequence is not an appropriate way to mine semantics. Despite the remarkable emergence ability of large language models~(LLMs) such as T5 and LLaMA2~\cite{zhao2023survey}, they have insignificant advantage in our method \wrt semantic modeling, and advanced strategies leveraged for LLM-empowered recommender systems still wait to be explored.

\section{Related work}
\label{sec:sec-related}

\paratitle{Sequential Recommendation.}
Sequential recommendation models leverage the chronological item sequences of users to understand their preferences and recommend following items. 
With the development of deep neural networks, various complicated model architectures have been leveraged to better characterize user preferences for sequential recommendation, including convolutional neural networks~\cite{tang2018caser} and recurrent neural networks~\cite{tan2016gru4rec, li2017narm}. In line with the emergence of the self-attention mechanism in Transformer~\cite{vaswani2017attention}, SASRec~\cite{kang2018sasrec}, BERT4Rec~\cite{sun2019bert4rec} and CL4SRec~\cite{xie2022contrastive} present better performance on extracting essential features.
Additionally, some of prior work take advantage of graph neural networks~(GNN) to capture high-order structures~\cite{wu2019session, xu2019graph, liu2020deoscillated}, involving item transition patterns in historical interaction sequences. 
Furthermore, Zhou \etal~\cite{zhou2022fmlp-rec} propose an all-MLP architecture with learnable filters to enhance recommendation performance. 
Latest work further develops the architecture of filter-based algorithms in the frequency domain to attenuate noise and explore periodic preferences~\cite{liu2023dlfs-rec, shin2023attentive, du2023fearec}.

\paratitle{Side Information Fusion.}
Side information fusion in sequential recommendation aims to integrate side information~(\eg item attributes) for better representing items and improving next-item predictions. 
Some prior work leverages late fusion and pre-training strategies based on self-attention mechanism to fuse side information, including FDSA~\cite{zhang2019fdsa} and $S^3$-Rec~\cite{zhou2020s3rec}. 
Recent studies concentrate on encoding side information with item representations as the input of recommendation models, 
including concatenation, graph-embedding aggregation and attention fusion~\cite{hidasi2016gru4recf, liu2021nova}. In addition to the early fusion of embedding, DIF-SR~\cite{xie2022dif-sr} moves the side information from the input to the attention layer and decouples the attention calculation for side information. Since items and their side information can be represented by
stochastic Gaussian distribution, DLFS-Rec~\cite{liu2023dlfs-rec} makes usage of mean and covariance embeddings of the distribution to formulate final embeddings for each item.

\paratitle{Language Models for Recommender Systems.}
Pre-trained language models~(PLMs) are proficient in natural language understanding and reasoning tasks, which can be categorized into two types: discriminative models and generative models. Existing literature utilizes discriminative models to generate textual representations of users and items for recommender systems~\cite{qiu2021u, zhang2022gbert, li2023text}. For example, UniSRec~\cite{hou2022unisrec} tends to combine various textual information~(\eg item descriptions) and designs adaptive model architecture to enhance the ability of universal representation learning. 
On the other hand, several researchers utilize generative language models and convert recommendation tasks to either language understanding or generation tasks~\cite{cui2022m6,geng2022p5}. Besides, generative language models can also be leveraged for data augmentation, \eg KAR~\cite{xi2023towards} and LLMRG~\cite{wang2023enhancing} extract abundant types of textual information and aggregate them into fused embeddings through MoE or GNN~\cite{xu2019graph}.

\begin{table}[t]
\small
\centering
\captionsetup{font={small}}
\caption{Performance comparison \wrt textual representations. }
\label{tab:lm-text-repre}
\begin{tabular}{@{}cccccc@{}}
\toprule
\multirow{2}{*}{LM}        & \multirow{2}{*}{Pooling} & \multicolumn{2}{c}{MovieLens-1M}  & \multicolumn{2}{c}{Online Retail} \\ \cmidrule(l){3-6} 
                           &                          & R@10            & N@10            & R@10            & N@10            \\ \midrule
Random                     & -                        & 0.2315          & 0.1238          & 0.1676          & 0.0865          \\ \midrule
\multirow{2}{*}{BERT-base} & CLS                      & \textbf{0.2623} & \textbf{0.1445} & \textbf{0.2234} & \textbf{0.1316} \\
                           & Mean                     & 0.2541          & 0.1402          & 0.2217          & 0.1310          \\ \midrule
T5-base                    & Mean                     & 0.2530          & 0.1403          & 0.2209          & 0.1299          \\ \midrule
Flan-T5-base               & Mean                     & 0.2546          & 0.1405          & 0.2220          & 0.1306          \\ \midrule
LLaMA2-7B                  & Mean                     & 0.2579          & 0.1420          & 0.2051          & 0.1176          \\ \bottomrule
\end{tabular}
\end{table}

\section{Conclusion}
In this paper, we propose a novel {\ul \emph{Te}}xt-I{\ul \emph{D}} semantic fusion approach for sequential {\ul \emph{Rec}}ommendation, namely \textbf{\our}. 
The core idea of our approach is to conduct a sequence-level semantic representation fusion approach by better integrating global contexts. 
To achieve this, we devise an MoE enhanced modulation method to improve the discriminability of textual representations from the text encoder, by adaptively injecting
positional information.
Furthermore, we transform the text and ID embeddings by Fourier Transform from  \emph{time domain} to \emph{frequency domain}, and employ multiplicative operations to effectively fuse the two kinds of item features. Our fusion approach is simple yet effective, which can be proved to have the same effects of contextual convolution for sequence-level semantic fusion. 
Extensive experiments conducted on five public datasets demonstrate the effectiveness and efficiency of our approach.
For future work, we will further consider leveraging the contextual convolution strategy in the frequency domain to more recommendation tasks~(\eg multi-modal recommendation) and other aspects of recommendation~(\eg diversified recommendation).


\bibliographystyle{ACM-Reference-Format}
\bibliography{reference}

\end{document}